\documentclass[showpacs,pra,onecolumn]{revtex4-1}
\usepackage{amssymb}
\usepackage{amsmath}
\usepackage{graphicx}
\usepackage{subfigure}
\usepackage{epstopdf}
\usepackage{color}

\begin{document}

\title{Multipoles and vortex multiplets in multidimensional media with
inhomogeneous defocusing nonlinearity}
\date{\today}
\author{Rodislav Driben$^{1,2,}$\thanks{%
Author to whom any correspondence should be addressed.}}
\author{Nir Dror$^{3}$}
\author{Boris A. Malomed$^{3}$}
\author{Torsten Meier$^1$}
\affiliation{$^1$Department of Physics \& CeOPP, University of
Paderborn, Warburger Str. 100,
Paderborn D-33098, Germany \\
$^2$ITMO University, 49 Kronverskii Ave., St. Petersburg 197101, Russian Federation\\
$^3$Department of Physical Electronics, School of Electrical
Engineering, Faculty of Engineering, Tel Aviv University, Tel Aviv
69978, Israel}


\begin{abstract}
We predict a variety of composite quiescent and spinning two- and
three-dimensional (2D and 3D) self-trapped modes in media with a repulsive
nonlinearity whose local strength grows from center to periphery. These are
2D dipoles and quadrupoles, and 3D octupoles, as well as vortex-antivortex
pairs and quadruplets. Unlike other multidimensional models, where such
complex bound states either do not exist or are subject to strong
instabilities, these modes are remarkably robust in the present setting. The
results are obtained by means of numerical methods and analytically, using
the Thomas-Fermi approximation. The predicted states may be realized in
optical and matter-wave media with controllable cubic nonlinearities.
\end{abstract}

\maketitle

\section{Introduction}

Spatial and spatiotemporal solitons in two- and three-dimensional (2D and
3D) settings are one of central research topics in photonics \cite{book}-%
\cite{RMP} and physics of quantum gases \cite{BEC,RMP}, as well as in other
areas, such as nuclear matter \cite{nuclear} and the field theory \cite%
{field-theory}. This topic also provides a strong incentive for studies of
soliton solutions and their stability in applied mathematics \cite{Yang}.
Typically, solitons result from the balance between diffraction and/or
group-velocity dispersion, which drive linear defocusing of the fields, and
their self-focusing induced by the attractive nonlinearity. While in one
dimension (1D) this mechanism readily gives rise to stable solitons \cite%
{book,RMP}, a fundamental problem in the 2D and 3D geometry is that the
conventional Kerr (cubic) nonlinearity leads to the critical (in 2D) and
supercritical (in 3D) collapse \cite{collapse}, hence the respective
multidimensional solitons are unstable.

Multipoles are a species of localized multidimensional modes which are of
significant interest to nonlinear optics \cite{MultipolesOptics} and studies
of matter waves in Bose-Einstein condensates (BEC) \cite{MultipolesBEC}, but
are also vulnerable to the collapse-induced instability. They appear as
multiple-peak solitons with alternating signs of adjacent peaks, the
simplest example being a dipole which consists of two lobes with opposite
signs of the field in them. While, as mentioned above, multipoles are
unstable in uniform Kerr media, possibilities were proposed to stabilize
them, including the introduction of optical lattices \cite{MultipolesLattice}%
, and the use of non-Kerr \cite{SaturableNonlinearity,propeller} and
nonlocal \cite{MultipolesNonlocal} nonlinearities.

Related to these are schemes elaborated for the stabilization of
multidimensional solitons in a more general context. One approach relies
upon the use of more complex photonic nonlinearities, such as saturable \cite%
{satur} (which are available in photorefractive crystals \cite{satur-photo}
and atomic vapors \cite{satur-vapor}), or cubic-quintic (CQ) \cite{CQ} and
quadratic-cubic \cite{QC,Buryak} combinations of competing focusing and
defocusing terms. Recently, the stability of 2D spatial solitons in optical
media featuring the CQ \cite{Cid} and quintic-septimal \cite{Cid2} nonlinear
responses has been demonstrated experimentally.

Nonlocal photonic nonlinearities also readily provide stabilization of
multidimensional solitons against small perturbations \cite{nonlocal}. The
solitons in nonlocal media are as well stable to thermal fluctuations \cite%
{fluct} and partial loss of incoherence \cite{incoh}. Peculiarities of the
modulational instability driven by the nonlocal nonlinearity, which is the
precursor of the formation of solitons, were studied too \cite{MI}.

The stabilization of 2D solitons may also be secured by \textquotedblleft
management" techniques, i.e., periodic alternation of the sign of the
nonlinear term \cite{management}. Still another possibility for the
stabilization is provided by trapping potentials -- in particular, periodic
lattices, as has been predicted theoretically for various settings \cite%
{lattice,SpatialSolitons,RMP} and was demonstrated experimentally in
photorefractive crystals \cite{lattice-experiment}.

A novel collapse-free scheme for the creation of fundamental and
topologically structured $D$-dimensional solitons was proposed in Ref. \cite%
{Barcelona} and further developed in a number of subsequent publications
\cite{others}-\cite{SciRep}. It is based on using the \emph{self-defocusing}
cubic nonlinearity with the local strength \emph{growing} at $r\rightarrow
\infty $ at any rate exceeding $r^{D}$, where $r$ is the distance from the
center. In reality, the local strength does not need to attain extremely
large values, as the solitons supported by this setting are strongly
localized, hence properties of the medium at distances essentially exceeding
the transverse size of the soliton are unimportant \cite{Barcelona}. In
optics, the 2D version of the scheme can be implemented for spatial solitons
by means of inhomogeneously doping the bulk waveguide by
nonlinearity-enhancing resonant impurities \cite{Kip}. In fact, a uniform
dopant density may be used, with an inhomogeneous distribution of detuning
from the two-photon resonance imposed by an external field \cite{Barcelona}.
In 2D and 3D atomic BECs, the spatial modulation of the scattering length of
two-body collisions, which determines the local nonlinearity strength in the
mean-field approximation, can be imposed by means of the optically
controlled spatially nonuniform Feshbach resonance, as experimentally
demonstrated in Ref. \cite{experiment-inhom-Feshbach}. Alternatively, the
required spatial profile of the nonlinearity can be ``painted", as an
averaged one, by a fast moving laser beam \cite{painting}. Another
experimentally demonstrated method makes use of the magnetically-controlled
Feshbach resonance in an appropriately shaped magnetic lattice, into which
the condensate is loaded \cite{magn-latt}.

Furthermore, the scheme for the creation of solitons in media with the
spatially growing nonlinearity was extended for 2D discrete solitons \cite%
{discrete}, as well as for 2D systems with the spatially modulated strength
of the long-range dipole-dipole repulsion \cite{Raymond}. This scheme gives
rise to a variety of sturdy multidimensional states, including 2D and 3D
vortex solitons \cite{Barcelona,SciRep} and more complex 3D modes,\textit{\
viz}., soliton gyroscopes \cite{gyroscope}, vertical vortex-antivortex
hybrids \cite{hybrid}, and vortex tori with intrinsic twist (``hopfions"),
which carry two topological numbers \cite{Yasha}. These self-trapped states
feature exceptional robustness in comparison with models of other types, as
some states, such as the vortex-antivortex hybrids \cite{hybrid}
and hopfions \cite{Yasha}, do not exist in other single-component models,
while multidimensional vortex solitons, and soliton gyroscopes, are stable
only in the present setting.

The objective of the present work is to expand the class of robust 2D and 3D
modes supported by the spatially growing repulsive nonlinearity, by adding
other obviously important soliton species, such as the above-mentioned
multipoles (2D dipoles and quadrupoles, and 3D octupoles), and multiplets
(bound states) built of 2D vortices and antivortices, including
vortex-antivortex pairs (VAPs) and quadruplets (VAQs). Spinning dipoles and
vortex-antivortex multiplets are introduced too. The consideration is
performed by means of numerical methods, in combination with the
Thomas-Fermi approximation (TFA), which makes it possible to predict basic
results in an analytical form.
In particular, we demonstrate that there is a bifurcation which destabilizes
2D dipoles, replacing them by stable VAPs, and another bifurcation, which
replaces isotropic 2D VAQs by stable anisotropic modes of the same type
(AVAQs). Thus, the present model offers the simplest setting in which stable
multipoles and vortex-antivortex multiplets can be constructed in the
multidimensional space. In this respect, it may be compared to field-theory
models, where soliton complexes are created in sophisticated multi-component
settings \cite{gauge,field-theory}.


\section{The governing equations}

The basic model is represented by the scaled $D$-dimensional
Gross-Pitaevskii equation (GPE) which governs the evolution of the
mean-field wave function, $\psi (\mathbf{r},t)$, in BEC:
\begin{equation}
i\frac{\partial \psi }{\partial t}=-\nabla ^{2}\psi +\sigma (\mathbf{r}%
)\left\vert \psi \right\vert ^{2}\psi .  \label{1}
\end{equation}%
Here $t$ is time, $\nabla ^{2}$ is the 3D or 2D Laplacian, and $\sigma (r)$
represents the spatially growing local strength of the repulsive
nonlinearity. Following Ref. \cite{Barcelona}, we here adopt the steep
modulation profile, $\sigma (r)=\exp (r^{2}/2r_{0}^{2})$, where we set $%
r_{0}\equiv 1$ by means of straightforward rescaling. As well as in Ref.
\cite{Barcelona}, milder profiles, characterized by $\sigma (r)$ growing as $%
r^{\alpha }$ (with $\alpha >D$, as said above) are possible too, but the
anti-Gaussian one makes it possible to present results in a compact form.
Dynamical invariants of Eq. (\ref{1}) are the norm, the angular momentum,
and the Hamiltonian:%
\begin{eqnarray}
N &=&\int \int \int \left\vert \psi \left( x,y,z,t\right) \right\vert
^{2}dxdydz,  \label{N} \\
\mathbf{M} &=&-i\int \int \int \psi ^{\ast }\left( \mathbf{r}\times \nabla
\right) \psi dxdydz,  \label{M} \\
H &=&\int \int \int \left[ \left\vert \nabla \psi \right\vert
^{2}+(1/2)\sigma (r)|\psi |^{4}\right] dxdydz.  \label{H}
\end{eqnarray}%
In the 2D reduction of the model, the vectorial angular momentum (\ref{M})
is replaced by its single component,%
\begin{equation}
M_{z}=i\int \int \psi ^{\ast }\left( y\frac{\partial }{\partial x}-x\frac{%
\partial }{\partial y}\right) \psi dxdy.  \label{Mz}
\end{equation}

The 2D version of Eq. (\ref{1}), with time substituted by the propagation
distance ($z$), applies as well to the spatial-domain evolution of the
amplitude of electromagnetic waves in a bulk optical waveguide \cite{book}
with the self-defocusing cubic nonlinearity. In that case, the 2D reduction
of Eq. (\ref{N}) determines the total power of the optical beam, while Eq. (%
\ref{Mz}) is the beam's orbital angular momentum \cite{OAM}.

Stationary states with real chemical potential $\mu >0$ (in optics, $-\mu $
is the propagation constant) are looked for in the usual form,
\begin{equation}
\psi \left( \mathbf{r},t\right) =\exp \left( -i\mu t\right) U(\mathbf{r}),
\label{U}
\end{equation}%
where the (generally, complex) spatial wave function obeys the equation%
\begin{equation}
\mu U=-\left( \frac{\partial ^{2}}{\partial r^{2}}+\frac{1}{r}\frac{\partial
}{\partial r}+\frac{1}{r^{2}}\frac{\partial ^{2}}{\partial \varphi ^{2}}%
\right) U+\exp \left( \frac{r^{2}}{2}\right) \left\vert U\right\vert ^{2}U
\label{2D}
\end{equation}%
in the 2D setting, with angular coordinate $\varphi $. The 3D stationary
equation is%
\begin{equation}
\mu U=-\left( \frac{\partial ^{2}}{\partial r^{2}}+\frac{2}{r}\frac{\partial
}{\partial r}+\frac{\mathbf{\hat{l}}^{2}}{r^{2}}\right) U+\exp \left( \frac{%
r^{2}}{2}\right) |U|^{2}U,  \label{3D}
\end{equation}%
where $\mathbf{\hat{l}}^{2}$\ is the usual angular part of the 3D Laplacian.

\section{Multipoles: Numerical results and the Thomas-Fermi approximation
(TFA)}

Stationary solutions of Eq. (\ref{2D}) and Eq. (\ref{3D}) for multipole
modes were obtained by means of the imaginary-time method \cite{IT}, which
is capable to generate not only the ground state but, under special
conditions, higher-order modes too \cite{IT-higher-modes}, and also by means
of the modified squared-operator method \cite{Lacoba}. The respective
inputs, emulating the multipoles sought for, were built as combinations of
identical Gaussians with separated centers and alternating signs. Typical
numerically generated examples of stationary multipoles, \textit{viz}., 2D
dipole and quadrupole, and a 3D octupole, are displayed in Figs. \ref%
{SolutionProfiles}(a), (b), and (c), respectively, for $\mu =2$.

The stability of the so generated modes was examined via direct simulations
of the perturbed evolution, using the standard pseudospectral split-step
fast-Fourier-transform method. 2D simulations were performed in the domain
of size $(12\pi )^{2}$, covered by a mesh of $512^{2}$ points. For the 3D
simulations of the octupoles, the split-step algorithm was used too, in the
domain of size $\left( \mathbf{12\pi }\right) ^{3}$, with the mesh of $%
192^{3}$ points. In particular, the course of the evolution, under the
influence of the small numerical noise due to the discrete nature of the
mesh, the octupole with $\mu =2$ kept its 3D shape virtually invariant for $%
t=200$, which exceeds 10 respective diffraction lengths.

\begin{figure}[t]
\subfigure[]{\includegraphics[width=2.2in]{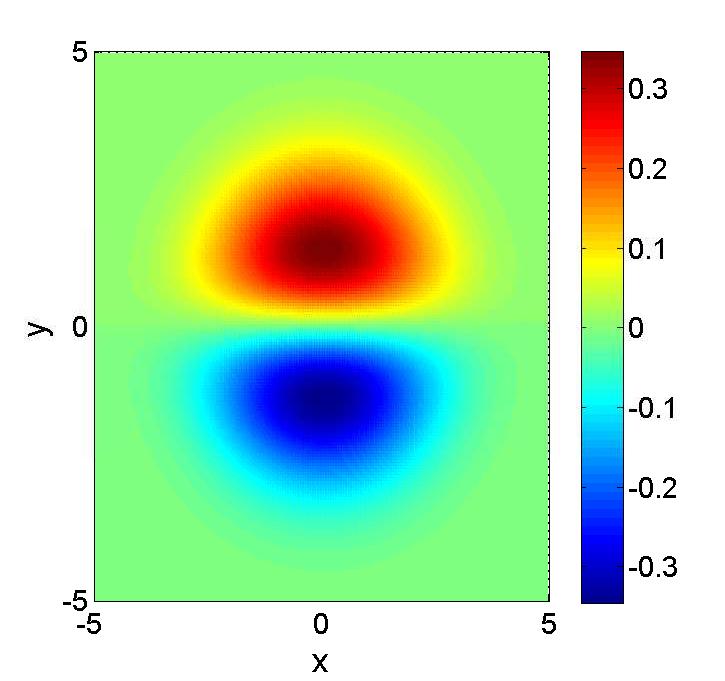}
\label{DipoleProfileMu5}}
\subfigure[]{\includegraphics[width=2.2in]{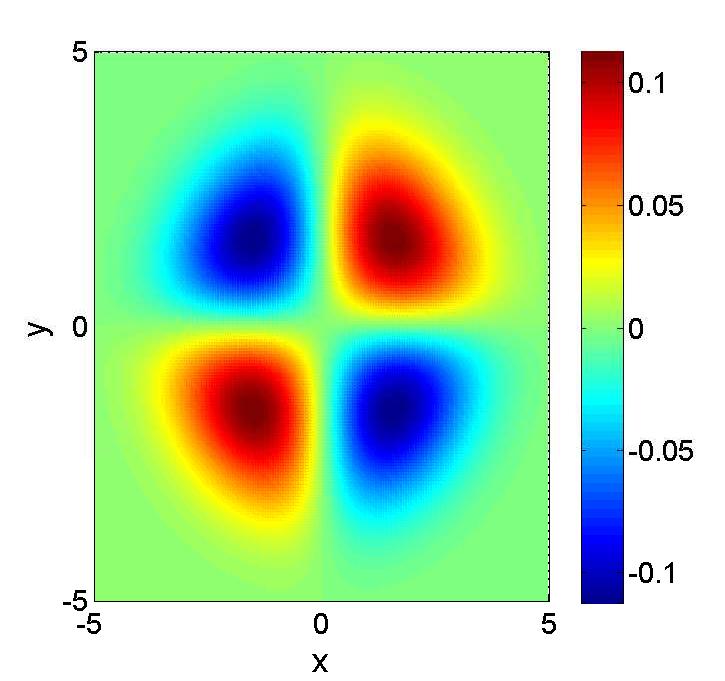}
\label{QuadrupoleProfileMu5}}
\subfigure[]{\includegraphics[width=2.2in]{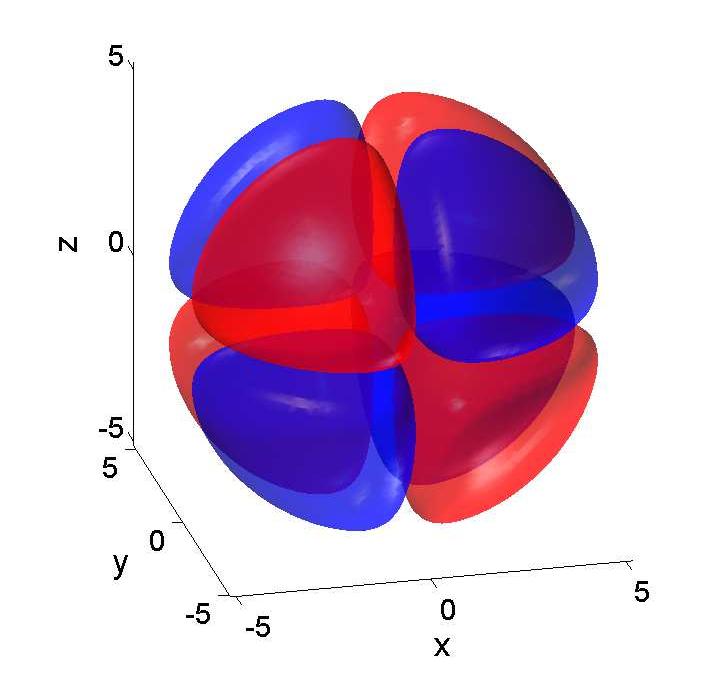}
\label{OctupoleProfile}}
\caption{(Color online) Generic examples of numerically found dipole (a) and
quadrupole (b) modes in the 2D setting, and of the octupole (c) in 3D. These
modes are described by real wave functions $U$ with $\protect\mu =2$, see
Eq. (\protect\ref{U}). In (c), the 3D image of the octupole is displayed by
means of red and blue isosurfaces, which correspond, respectively, to $%
U=+0.01$ and $U=-0.01$.}
\label{SolutionProfiles}
\end{figure}

Numerically found shapes of the 2D dipole ($m=1$) and quadrupole ($m=2$)
modes, which are displayed in Figs. \ref{SolutionProfiles}(a) and (b),
suggest that these modes may be approximated, in the simplest form, by a
real \textit{ansatz},%
\begin{equation}
U_{m}(r,\varphi )=V_{\mathrm{2D}}(r)\cos \left( m\varphi \right) .
\label{UV}
\end{equation}%
The substitution of this into Eq. (\ref{UV}), and the use of the usual
mean-field approximation, $\cos ^{3}\left( m\varphi \right) \rightarrow
(3/4)\cos \left( m\varphi \right) $, yields the following radial equation:%
\begin{equation}
\mu V_{\mathrm{2D}}=-\left( \frac{d^{2}}{dr^{2}}+\frac{1}{r}\frac{d}{dr}-%
\frac{m^{2}}{r^{2}}\right) V_{\mathrm{2D}}+\frac{3}{4}\exp \left( \frac{r^{2}%
}{2}\right) V_{\mathrm{2D}}^{3}.  \label{V}
\end{equation}%
The application of the TFA to Eq. (\ref{V}) implies neglecting the
derivative terms, which yields the simple result:%
\begin{equation}
V_{\mathrm{2D}}^{2}(r)=\left\{
\begin{array}{c}
0,~r^{2}<\frac{m^{2}}{\mu }, \\
\frac{4}{3}\left( \mu -\frac{m^{2}}{r^{2}}\right) \exp \left( -\frac{r^{2}}{2%
}\right) ,~r^{2}>\frac{m^{2}}{\mu }~.%
\end{array}%
\right.  \label{TF}
\end{equation}

This approximation predicts that maxima of the amplitude are located at
distance from the origin%
\begin{equation}
r_{\mathrm{2D}}^{(\max )}=\sqrt{\frac{|m|}{2\mu }\left( \sqrt{8\mu +m^{2}}%
+|m|\right) }.  \label{max}
\end{equation}%
{\LARGE \ } Amplitude maxima of the numerical solutions, found for $\mu =2$,
are located at distances $r_{\mathrm{dipole}}^{(\max )}=1.37$, $r_{\mathrm{%
quadrupole}}^{(\max )}=2.2$, and $r_{\mathrm{octupole}}^{(\max )}=3.46$ from
the origin, while the TFA results (\ref{max}) and (\ref{max-3D}) (see below)
predict, for the same case ($\mu =2$), $r_{\mathrm{dipole}}^{(\max )}=1.13$,
$r_{\mathrm{quadrupole}}^{(\max )}=1.8$, and $r_{\mathrm{octupole}}^{(\max
)}=2.75$. Naturally, the agreement improves for larger $\mu $, i.e., larger $%
N$, as the TFA is more accurate for stronger nonlinearity (the increase of $%
\mu $ from $2$ to $10$ leads to a decrease of the relative difference
between the numerical solutions and their TFA counterparts by a factor $%
\simeq 2$).

Families of the modes are characterized by the norm (total power) as a
function of $\mu $. The TFA based on Eq. (\ref{TF}) produces the following
dependence:%
\begin{equation}
N_{\mathrm{2D}}^{\left( \mathrm{TFA}\right) }(\mu ;m)=\frac{2\pi }{3}%
\int_{m^{2}/\left( 2\mu \right) }^{\infty }\left( 2\mu -\frac{m^{2}}{\rho }%
\right) e^{-\rho }d\rho ,  \label{NTF}
\end{equation}%
where $\rho \equiv r^{2}/2$, and the integral can be easily computed
numerically. In the limit of $\mu \rightarrow \infty $, Eq. (\ref{NTF})
yields an asymptotically constant slope,%
\begin{equation}
\frac{d}{d\mu }N_{\mathrm{2D}}^{\left( \mathrm{TFA}\right) }\left( \mu
\rightarrow \infty \right) =\frac{4\pi }{3}.  \label{dN/dmu}
\end{equation}%
Dependences (\ref{NTF}) for $m=1$ and $m=2$ are displayed in Fig. \ref%
{NvsMuDipoleQuadrupoleOctupole}(a), along with their counterparts produced
by full numerical solutions of Eq. (\ref{2D}).

\begin{figure}[t]
\subfigure[]{\includegraphics[width=2.5in]{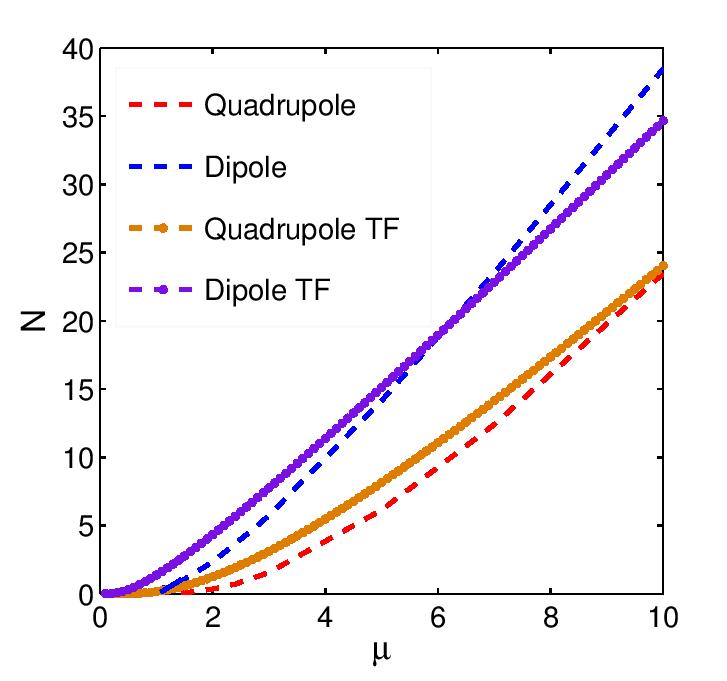}
\label{NvsMuDipoleQuadrupole}}
\subfigure[]{\includegraphics[width=2.5in]{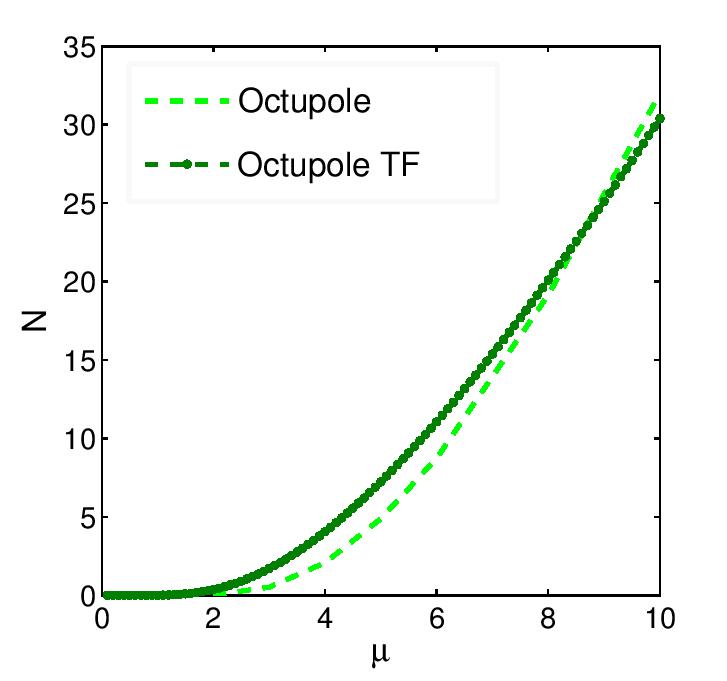}
\label{NvsMuOctupole}}
\caption{(Color online) Lines labeled \textquotedblleft Dipole",
\textquotedblleft Quadrupole", and \textquotedblleft Octupole" show,
respectively, the 2D and 3D norms, $N$, vs. $\protect\mu $ for these modes,
as obtained from numerical computations. The lines marked by
\textquotedblleft TF" display the predictions for the same dependences
produced by the Thomas-Fermi approximation, as per Eqs. (\protect\ref{NTF})
and (\protect\ref{NTF-3D}). The short solid segment of the $N(\protect\mu )$
curve for the dipole-mode family designates a completely stable subfamily
(cf. Fig. \protect\ref{fig5} below).}
\label{NvsMuDipoleQuadrupoleOctupole}
\end{figure}

In the 3D geometry, the typical shape of the octupoles, displayed in Fig. %
\ref{SolutionProfiles}(c), suggests that the corresponding angular structure
may be approximated by spherical harmonics with quantum numbers $l=3,m=\pm 2$
\cite{LL}: $Y_{32}\equiv Y_{3}^{+2}+Y_{3}^{-2}=C\cos \vartheta \sin
^{2}\vartheta \cos \left( 2\varphi \right) $, where the constant $C$ is not
essential here, $\vartheta $\ is the second angular coordinate, and $\mathbf{%
\hat{l}}^{2}Y_{32}=12Y_{32}$. Thus, the ansatz for the octupole is adopted
as
\begin{equation}
U(r,\vartheta ,\varphi )=V_{\mathrm{3D}}(r)\cos \vartheta \sin ^{2}\vartheta
\cos \left( 2\varphi \right) .  \label{3Dansatz}
\end{equation}%
For the application of the mean-field approximation to the angular
dependence in the cubic term of Eq. (\ref{3D}), $\left[ Y_{32}\left(
\vartheta ,\varphi \right) \right] ^{3}$ must be projected back onto $%
Y_{32}\left( \vartheta ,\varphi \right) $, which is done with the help of
the standard formula, $\left\langle f(\vartheta )\right\vert \left\vert
g(\vartheta )\right\rangle =\int_{0}^{\pi }f(\vartheta )g(\vartheta )\left(
\sin \vartheta \right) d\vartheta $. The eventual results produced by the
3D\ version of the TFA are%
\begin{equation}
V_{\mathrm{3D}}^{2}(r)\approx \left\{
\begin{array}{c}
0,~r^{2}<\frac{12}{\mu }, \\
\frac{143}{12}\left( \mu -\frac{12}{r^{2}}\right) \exp \left( -\frac{r^{2}}{2%
}\right) ,~r^{2}>\frac{12}{\mu }~.%
\end{array}%
\right.   \label{TF-3D}
\end{equation}%
cf. Eq. (\ref{TF}),
\begin{equation}
r_{\mathrm{3D}}^{(\max )}=\sqrt{\frac{6}{\mu }\left( \sqrt{\frac{2}{3}\mu +1}%
+1\right) },  \label{max-3D}
\end{equation}%
cf. Eq. (\ref{max}), and%
\begin{equation}
N_{\mathrm{3D}}^{\left( \mathrm{TFA}\right) }(\mu )=\frac{572\pi }{315}\int_{%
\sqrt{12/\mu }}^{\infty }\left( \mu r^{2}-12\right) e^{-r^{2}/2}dr,
\label{NTF-3D}
\end{equation}%
cf. Eq. (\ref{NTF}), with%
\begin{equation}
\frac{d}{d\mu }N_{\mathrm{3D}}^{\left( \mathrm{TFA}\right) }\left( \mu
\rightarrow \infty \right) =\frac{143}{315}(2\pi )^{3/2},  \label{dN/dmu3}
\end{equation}%
cf. Eq. (\ref{dN/dmu}). The plot produced by Eq. (\ref{NTF-3D}) is displayed
in Fig. \ref{NvsMuDipoleQuadrupoleOctupole}(b), along with its counterpart
generated by the full numerical solution.

An essential finding revealed by the systematic numerical analysis of the 2D
setting is that the increase of $\mu $ (and $N$) leads to destabilization of
the 2D dipole mode at a critical point,%
\begin{equation}
\mu _{\mathrm{cr}}^{\mathrm{(VAP)}}=1.64,~N_{\mathrm{cr}}^{\mathrm{(VAP)}%
}=1.38,  \label{cr}
\end{equation}%
by a \textit{bifurcation}, which gives rise to a new mode in the form of a
VAP, which will be presented in the next section. Further, careful analysis
demonstrates that both 2D quadrupoles and 3D octupoles are, strictly
speaking, entirely unstable solutions. Nevertheless, they are robust
objects, provided that their norms are relatively small, as the instability
is very weak in that case. For the quadrupoles, the simulations reveal that
the robustness interval is $\mu <\mu _{\max }\simeq 3$, corresponding to $%
N<N_{\max }\simeq 1.55$, see Fig. \ref{NvsMuDipoleQuadrupoleOctupole}(a). In
this interval, initially perturbed quadrupoles maintain their shape as long
as the simulations were run (with the evolution times measured in many
dozens of characteristic dispersion times), exhibiting small-amplitude
oscillations between the unperturbed shape and a coexisting stable isotropic
VAQ mode (the latter one is presented in detail below). For the octupoles a
similarly defined robustness border is located at $\mu _{\max }\simeq 2$,
corresponding to a low value of the 3D norm, $N\simeq 0.03$.

The application of the torque to the stable 2D dipole mode, i.e., the
multiplication of the respective stationary solution by the phase factor%
\begin{equation}
T=\exp \left( ipy\tanh \left( x/x_{0}\right) \right) ,  \label{torque}
\end{equation}%
with real constants $p$ and $x_{0}$ \cite{gyroscope}, readily sets it in
persistent rotation, thus creating a robust \textquotedblleft propeller"
mode, cf. Ref. \cite{propeller}, which preserves the original two-peak
shape. The spinning regime is illustrated by a set of snapshots in Fig. \ref%
{fig3}, where the rotation period is $T=3.9$. Rotation of the dipole was
simulated up to $t=100$, amplitude oscillations of the dipole in the course
of rotation being limited to $\simeq 2\%$ of its initial value.\textbf{\ }On
the other hand, the application of the torque to the quadrupoles and
octupoles does not generate persistently rotating states, in accordance with
the above conclusion that these species do not represent fully stable modes.

\begin{figure}[t]
\subfigure[]{\includegraphics[width=1.6in]{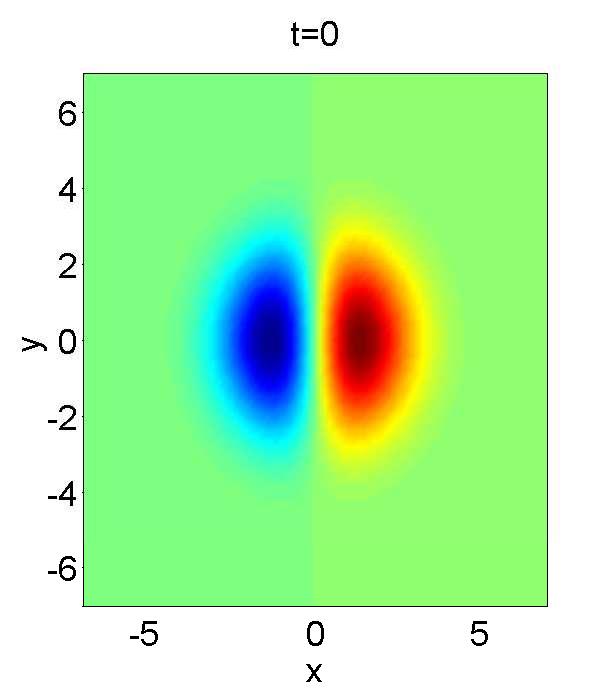}
\label{RotationDipoleMU2t0}}
\subfigure[]{\includegraphics[width=1.6in]{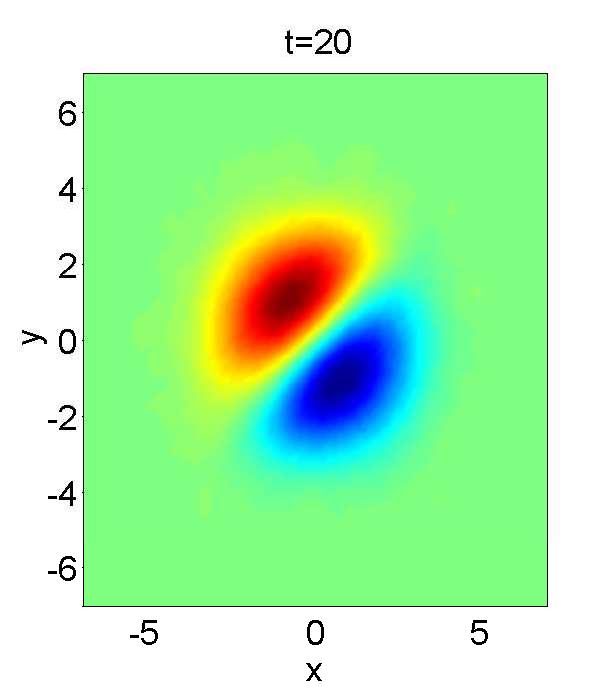}
\label{RotationDipoleMU2t20}}
\subfigure[]{\includegraphics[width=1.6in]{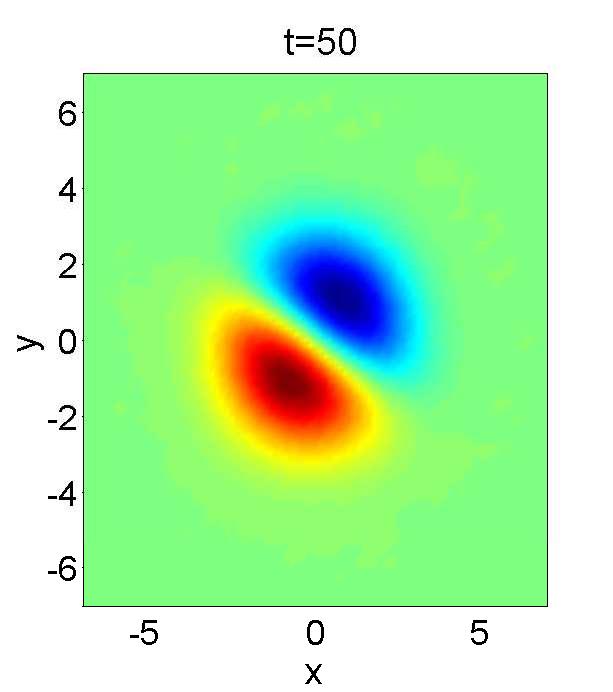}
\label{RotationDipoleMU2t50}}
\subfigure[]{\includegraphics[width=1.93in]{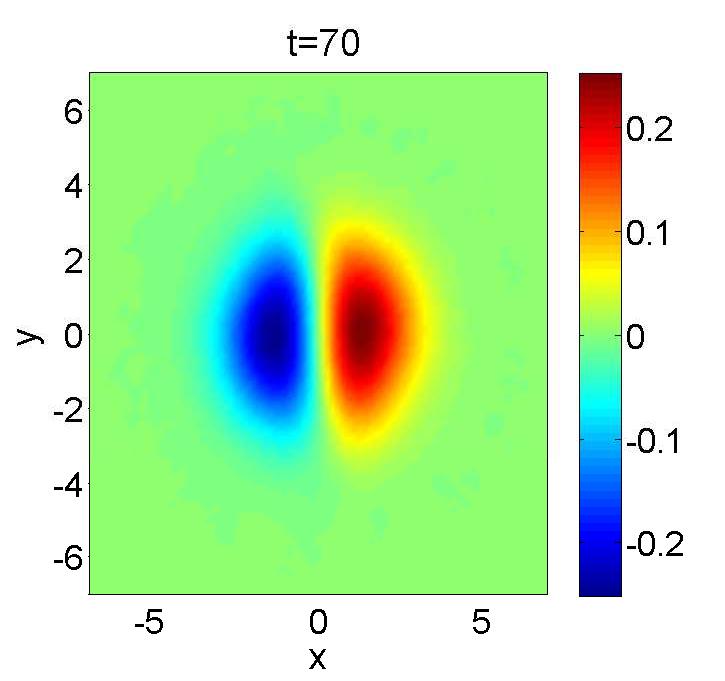}
\label{RotationDipoleMU2t70}}
\caption{Persistent rotation of a stable dipole with $\protect\mu =1.5$,
initiated by the application of a torque, given by Eq.~(\protect\ref{torque}%
) with $p=1$, $x_{0}=10$.}
\label{fig3}
\end{figure}

\section{Vortex-antivortex pairs (VAPs)}

\begin{figure}[tbp]
\subfigure[]{\includegraphics[width=2.4in]{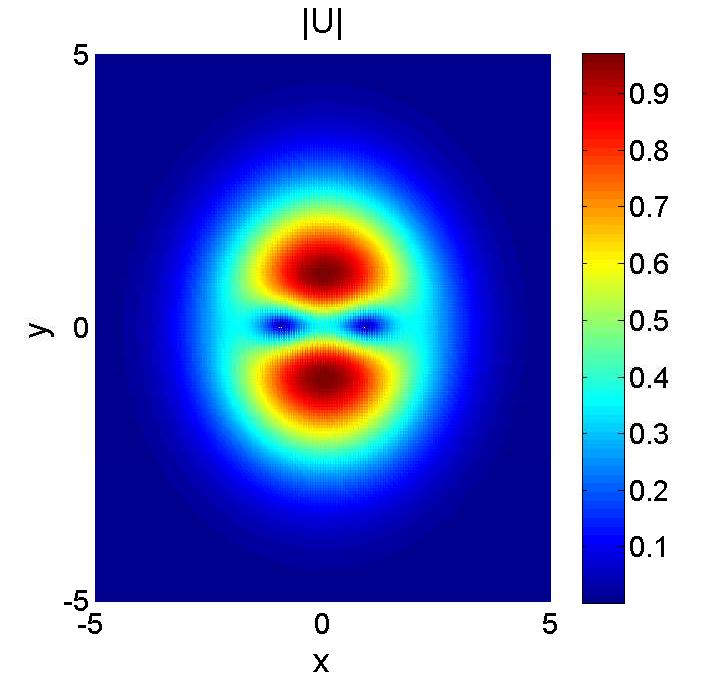}
\label{VortexPairAmpMu5}}
\subfigure[]{\includegraphics[width=2.4in]{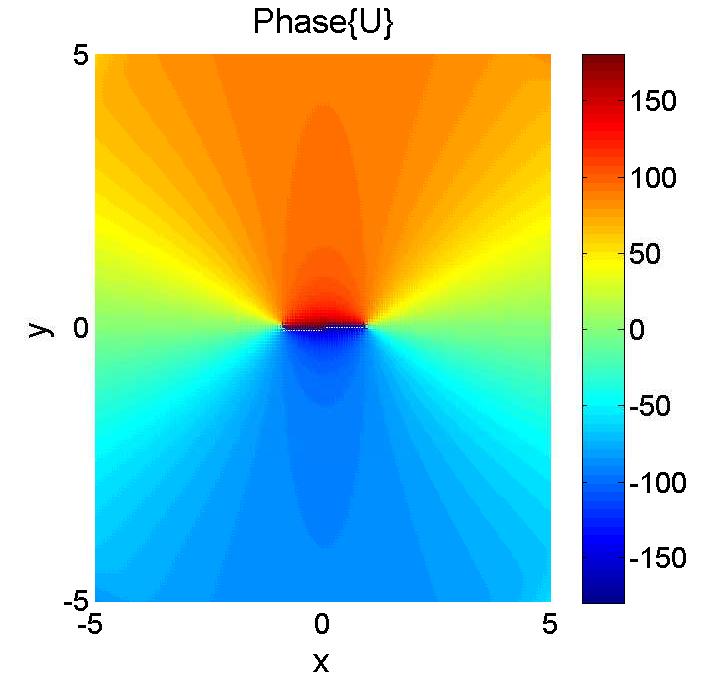}
\label{VortexPairPhaseMu5}}
\caption{An example of the amplitude and phase profiles of a 2D stable
vortex-antivortex pair (VAP), obtained in the numerical form for $\protect%
\mu =5$ (the corresponding norm is $N=14.78$). \ The phase distribution is
displayed in degrees.}
\label{IsotropicVAQ}
\end{figure}

\begin{figure}[t]
\centering\centering%
\includegraphics[width=2.4in]{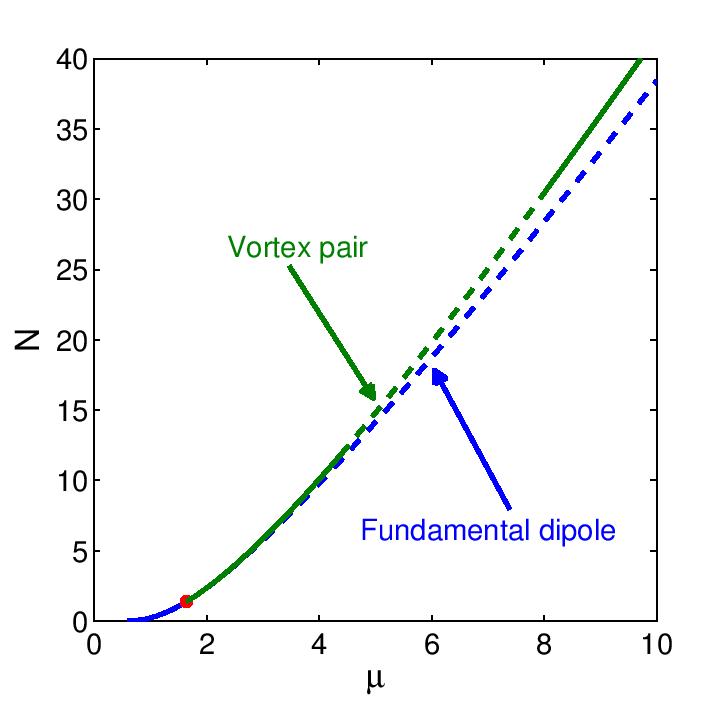}
\caption{$N(\protect\mu )$ curves for the dipole-mode and
vortex--antivortex-pair (VAP)\ families. Stable and unstable subfamilies are
indicated by continuous and dotted lines, respectively. The red dot marks
the bifurcation point (\protect\ref{cr}), at which the VAP family branches
off from the dipole-mode one.}
\label{fig5}
\end{figure}

In addition to multipoles, bound pairs of vortices with opposite signs of
angular momenta were investigated. A typical example of the VAP and the
respective bifurcation diagram are displayed in Figs. \ref{IsotropicVAQ} and %
\ref{fig5}, respectively. As seen in panel (a) of Fig. \ref{IsotropicVAQ},
the amplitude profile of the VAP is quite similar to that of the dipole
mode, but panel (b) of the same figure demonstrates the phase structure of
the stationary VAP, which its real dipole-mode counterpart does not have. It
is exactly the phase structure which makes it possible to identify the new
mode as the vortex-antivortex bound state.

In accordance with what is said above, the stability of the modes under the
consideration was identified by means of systematic simulations of their
perturbed evolution. As shown in Fig. \ref{fig5}, the VAP is stable after
the bifurcation point, \textit{viz}., at $1.64<\mu <4.06$, $1.38<N<10.35$,
then it is destabilized by oscillatory (in $t$) perturbations in a finite
region adjacent to the stability area, $4.06<\mu <8.06$, $10.35<N<30.76$,
which is followed by the eventual restabilization of the VAP at $\mu >8.06$,
$N>30.76$. Stable VAPs can be easily set in spinning motion, similar to what
is shown above for dipoles.

The emergence of the VAPs determines the instability of the dipole modes. In
the interval adjacent to the critical point Eq.~(\ref{cr}), $1.64<\mu <3$,
the dipole features a weak instability, periodically oscillating between its
unperturbed shape and the coexisting stable VAP, whose amplitude profile is
close to the dipole's one (not shown in detail here). 
A similar dynamical regime, featuring remittent shape revivals of a weakly
unstable dipole state, was observed in a model with a nonlocal nonlinearity
\cite{MultipolesNonlocal}. At $\mu >3$, the dipole's instability gets
stronger, leading to its spontaneous transformation into a fundamental
isotropic state (not shown in detail either).

\section{Vortex-antivortex quadruplets (VAQs)}

\label{sec:QuadrupoleVAQ}

\begin{figure}[tbp]
\subfigure[]{\includegraphics[width=2.4in]{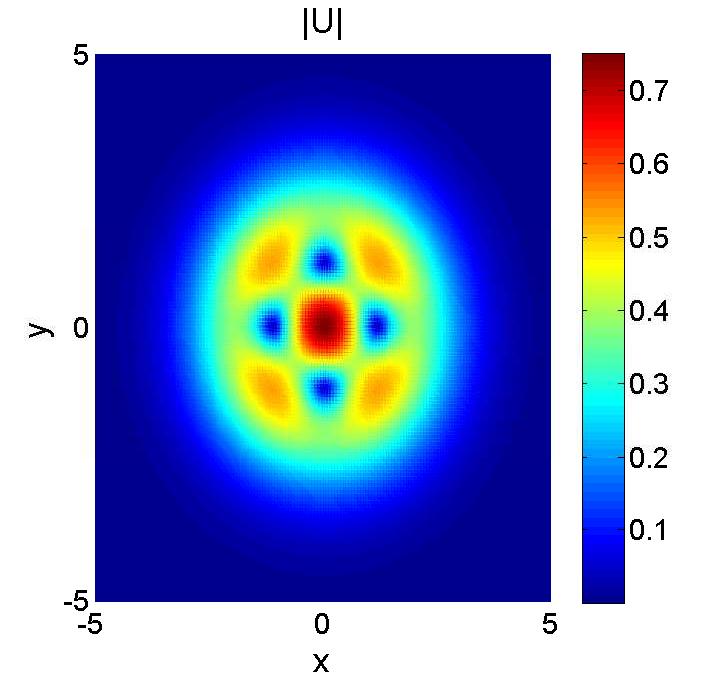}
\label{IsotropicVAQAmpMu5}}
\subfigure[]{\includegraphics[width=2.4in]{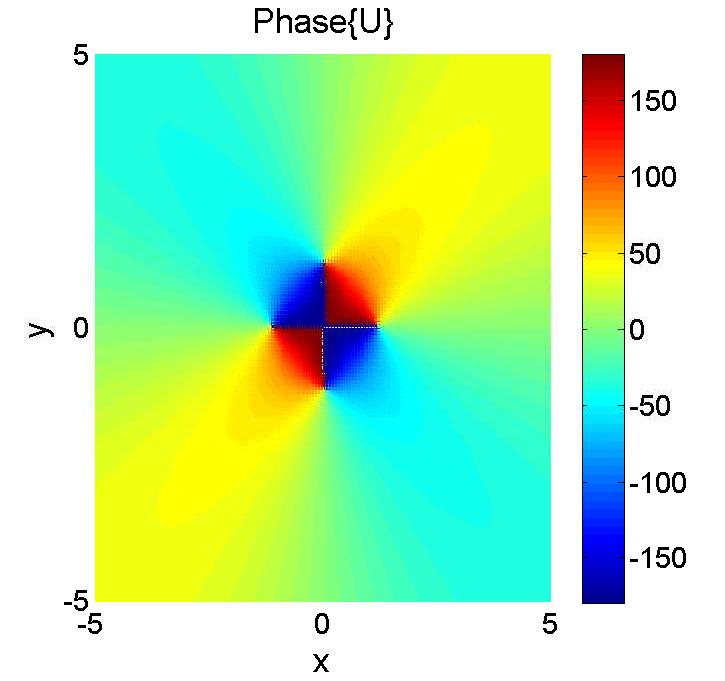}
\label{IsotropicVAQPhaseMu5}}
\caption{The amplitude (a) and phase (b) profiles of a stable isotropic
vortex-antovortex quadruplet (VAQ) at $\protect\mu =5$ and $N=8.12$.}
\label{IsotropicVAQProfile}
\end{figure}

\begin{figure}[tbp]
\includegraphics[width=2.4in]{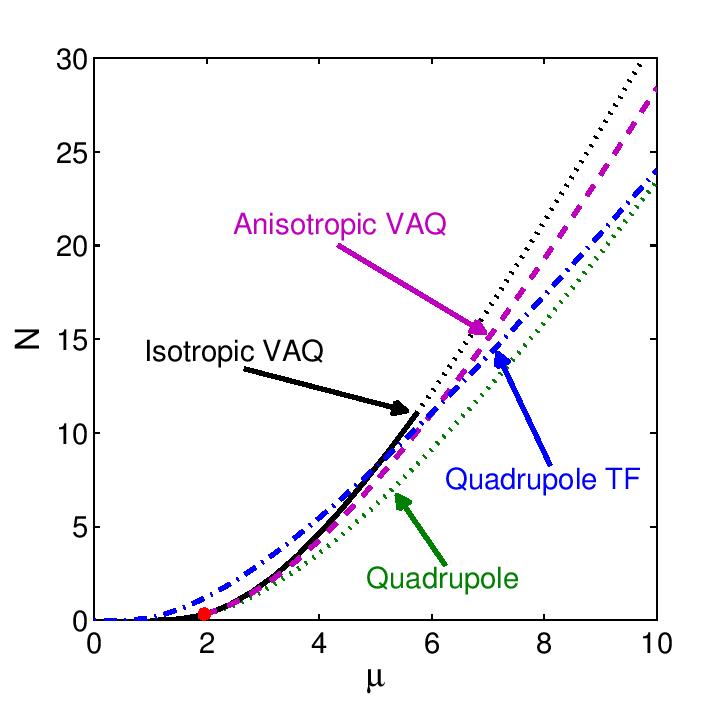}
\caption{Juxtaposed $N(\protect\mu )$ curves for the (fully unstable)
quadrupoles, as produced by the numerical results and by the Thomas-Fermi
approximation (\textquotedblleft TF") [these two curves are the same as in
Fig. \protect\ref{NvsMuDipoleQuadrupoleOctupole}(a)], and for the (partly
stable) isotropic and (fully unstable) anisotropic vortex-antivortex
quadruplets (VAQs). As usual, stable and unstable families of solutions are
indicated by solid and dotted lines, respectively. The dashed-dotted and
dashed lines refer, severally, to the Thomas-Fermi approximation for the
quadrupoles and the anisotropic VAQs. The bifurcation point (\protect\ref%
{aniso}), at which the anisotropic-VAQ branch splits off from its isotropic
counterpart, is marked by the red dot.}
\label{QuadrupoleVAQNvsMu}
\end{figure}

Similar to the 2D dipoles, which coexist with VAPs, real stationary 2D
quadrupoles coexist with a branch of \emph{stable} complex solutions for
vortex-antivortex bound states in the form of the (\textit{isotropic}) VAQ,
see a typical example of the latter in Fig. \ref{IsotropicVAQProfile}.
However, Fig. \ref{QuadrupoleVAQNvsMu} demonstrates the difference from the
situation for the coexistence of the dipoles and VAPs: the isotropic-VAQ
branch emerges at $\mu =0$, rather than branching off from the quadrupole
one at a finite value of $\mu $, cf. Fig. \ref{fig5}. Recall that, unlike
the dipoles, quadrupole solutions are, strictly speaking, always unstable,
hence, indeed, a new stable branch cannot bifurcate from them. As shown in
Fig. \ref{QuadrupoleVAQNvsMu} , the isotropic-VAQ family is stable at $0<\mu
<5.74$ (which corresponds to $0<N<11.06$). At $\mu >5.74$ ($N>11.06$), these
modes become unstable against oscillatory perturbations, eventually evolving
into the isotropic ground state (not shown here in detail).

%
%

\begin{figure}[tbp]
\subfigure[]{\includegraphics[width=2.4in]{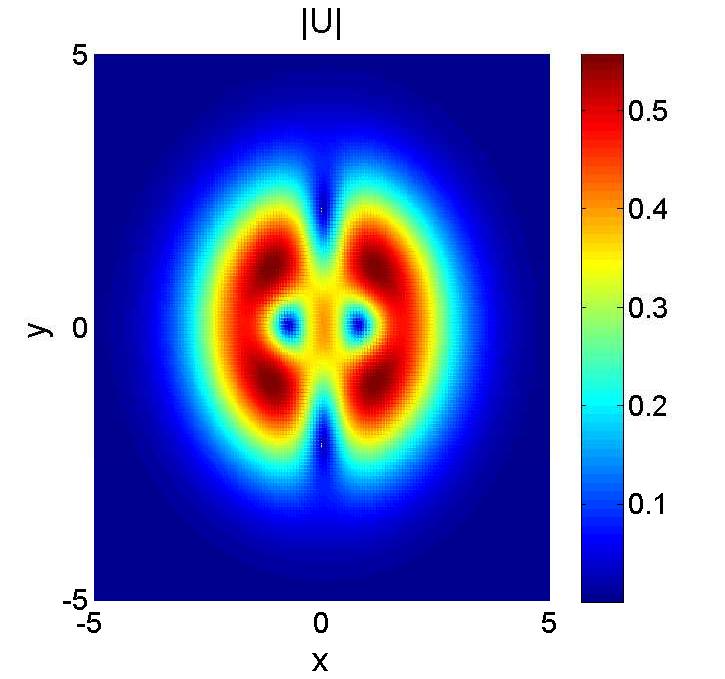}
\label{AnisotropicVAQAmpMu5}}
\subfigure[]{\includegraphics[width=2.4in]{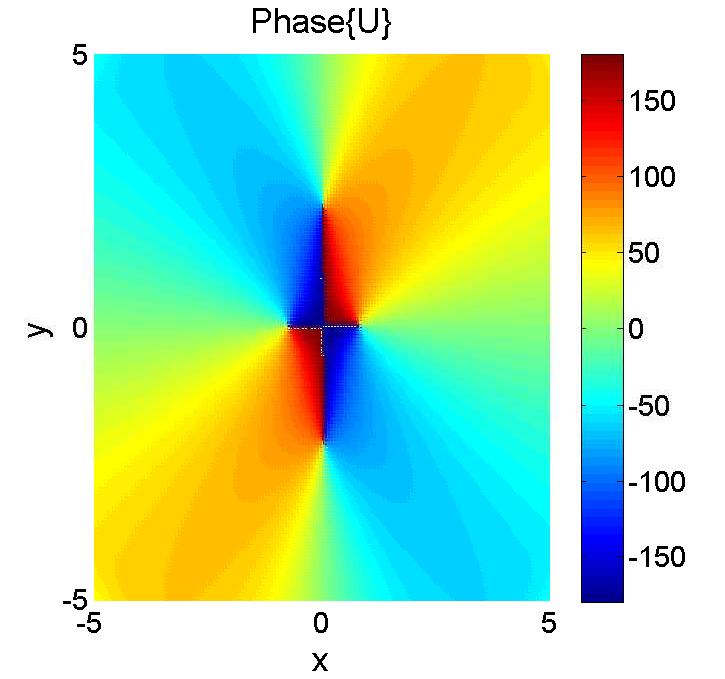}
\label{AnisotropicVAQPhaseMu5}}
\caption{The same as in Fig. \protect\ref{IsotropicVAQProfile}, but for an
unstable anisotropic VAQ, with $\protect\mu =5$, $N=7.40$.}
\label{AnisotropicVAQ}
\end{figure}

Further, a new branch, of \textit{anisotropic} VAQs (\textit{AVAQs}),
bifurcates from the family of their isotropic counterparts at
\begin{equation}
\mu _{\mathrm{cr}}^{(\mathrm{aniso})}=1.96,~N_{\mathrm{cr}}^{(\mathrm{aniso}%
)}=0.33.  \label{aniso}
\end{equation}%
A typical example of an AVAQ mode is displayed in Fig. \ref{AnisotropicVAQ}.
The analysis demonstrates that this solution is always unstable [this fact
explains why the bifurcation occurring at point (\ref{aniso}) does not
destabilize the branch of the isotropic VAQs]. Specifically, from the
bifurcation point (\ref{aniso}) up to $\mu =12.54$ ($N=40.77$), a perturbed
AVAQ solution starts spinning motion, with an angular velocity determined by
the initial perturbation, see an example in Fig. \ref%
{AnisotropicVAQEvolution} (in this case, the conservation of the angular
momentum is provided by the recoil effect of small-amplitude waves emitted
by the perturbed AVAQ, which are not visible in Fig. \ref%
{AnisotropicVAQEvolution}). Eventually, the unstable spinning AVAQ
transforms not into the isotropic ground state, which is typical for other
species of unstable modes which were mentioned above, but, instead, into a
stably rotating VAP. In several cases that were examined, this transition
occurred before the AVAQ would complete half a cycle of its rotation.
Finally, at $\mu >12.54$ ($N>40.77$), the AVAQ immediately transforms into a
rotating VAP, without going through the intermediate spinning stage (not
shown here in detail).



\begin{figure}[tbp]
\subfigure[]{\includegraphics[width=1.6in]{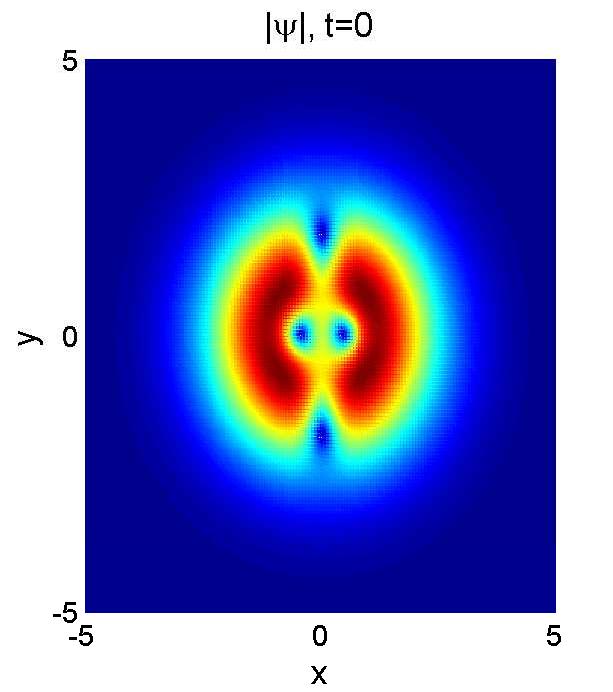}
\label{AnisotropicVAQEvolutionMu12t0}}
\subfigure[]{\includegraphics[width=1.6in]{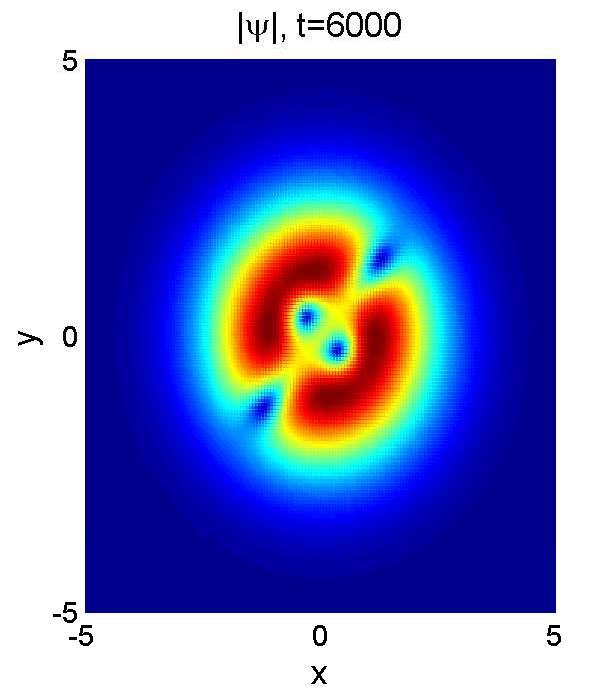}
\label{AnisotropicVAQEvolutionMu12t5000}}
\subfigure[]{\includegraphics[width=1.93in]{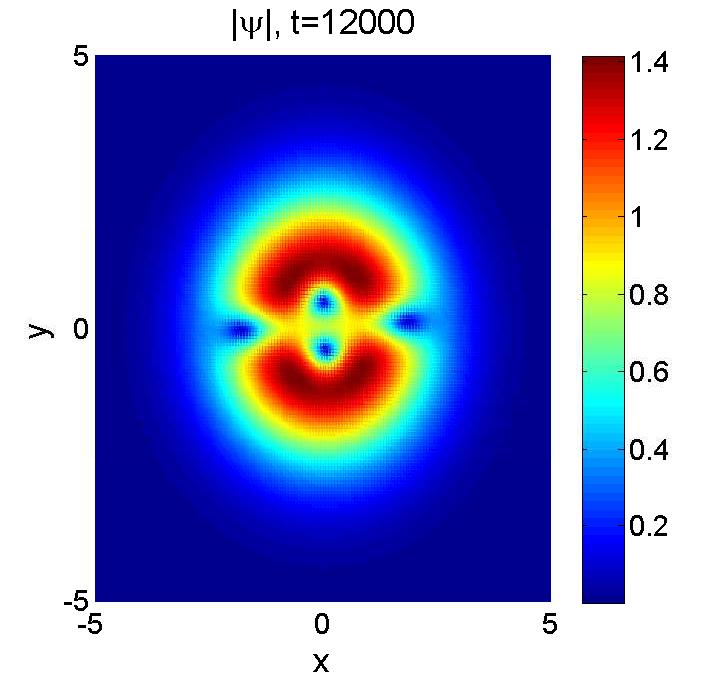}
\label{AnisotropicVAQEvolutionMu12t10000}}
\caption{The transient rotation of the unstable AVAQ, with $\protect\mu =12$%
, $N=38.08$, prior to its spontaneous transformation into a stable
vortex-antivortex pair. }
\label{AnisotropicVAQEvolution}
\end{figure}


\section{Conclusion}

The recently introduced class of optical and matter-wave models, with the
strength of the self-repulsive cubic nonlinearity growing from the center to
periphery, gives rise to a large variety of completely stable complex 2D and
3D self-trapped states (solitons), which do not exist or are strongly
unstable in other physical settings. Here we have demonstrated that,
together with previously found states, the same models support several other
species of 2D and 3D solitons which are not available (in a stable form) in
other systems either. These are 2D dipoles and quadrupoles and 3D octupoles,
as well as VAPs (vortex-antivortex pairs) and isotropic or anisotropic VAQs
(vortex-antivortex quadruplets). The modes found here are stable or weakly
unstable, surviving for a long time (over a long propagation distance),
which makes them physically relevant objects. In addition to quiescent
states, persistently spinning dipoles have been found too. The results,
which were obtained by means of numerical methods and analytically, using
the TFA\ (Thomas-Fermi approximation), demonstrate the potential of the
settings with the spatially growing strength of the self-repulsion for the
creation of a great variety of complex stable \ multidimensional modes.

Natural extensions of the present analysis may be additional analysis of the
3D setting (in particular, for 3D dipoles and quadrupoles), and its
development for two-component systems, in which the multitude of nontrivial
self-trapped states may be further expanded, as suggested, e.g., by results
for the two-component system in 1D \cite{two-component}.


\begin{acknowledgements}
RD and TM acknowledge support by the Deutsche Forschungsgemeinschaft (DFG) via the
Research Training Group (GRK) 1464 and computing time provided by PC$^2$
(Paderborn Center for Parallel Computing). RD acknowledges support by the Russian
Federation Grant 074-U01 through ITMO Early Career Fellowship scheme.
\end{acknowledgements}

\end{document}